\documentclass[aps,twocolumn,a4paper,nofootinbib,superscriptaddress]{revtex4}
\usepackage[dvips]{graphicx}
\usepackage{amsmath}
\usepackage{amssymb}

\begin{document}

\title{Edge magnetoplasmons in graphene: Effects of gate screening and dissipation}
\author{Alexey A. Sokolik}%
\affiliation{Institute for Spectroscopy, Russian Academy of Sciences, 142190 Troitsk, Moscow, Russia}%
\affiliation{National Research University Higher School of Economics, 109028 Moscow, Russia}%
\author{Yurii E. Lozovik}%
\email{lozovik@isan.troitsk.ru}
\affiliation{Institute for Spectroscopy, Russian Academy of Sciences, 142190 Troitsk, Moscow, Russia}%
\affiliation{National Research University Higher School of Economics, 109028 Moscow, Russia}%
\affiliation{Dukhov Research Institute of Automatics (VNIIA), 127055 Moscow, Russia}%

\begin{abstract}
Magnetoplasmons on graphene edge in quantizing magnetic field are investigated at different Landau level filling
factors. To find the mode frequency, the optical conductivity tensor of disordered graphene in magnetic field is
calculated in the self-consistent Born approximation, and the nonlocal electromagnetic problem is solved using the
Wiener-Hopf method. Magnetoplasmon dispersion relations, velocities and attenuation lengths are studied numerically and
analytically with taking into account the screening by metallic gate and the energy dissipation in graphene. The
magnetoplasmon velocity decreases in the presence of nearby gate and oscillates as a function of the filling factor
because of the dissipation induced frequency suppression occurring when the Fermi level is located near the centers of
Landau levels, in agreement with the recent experiments.
\end{abstract}

\maketitle

\section{Introduction}
Two-dimensional plasmons on graphene offer ample opportunities of applications due to wide tunability of their
properties achieved by changing the doping level, confining charge carriers, by nanostructuring  graphene or combining
it with metal electrodes \cite{Politano,Xiao,Goncalves}. In magnetic field the plasmon resonance splits into two
magnetoplasmon modes, as found using the terahertz spectroscopy of graphene disks \cite{Crassee,Yan}. The
higher-frequency mode can be treated in the quasiclassical limit as two-dimensional plasma oscillations acquiring a
frequency enhancement due to confining action of magnetic field \cite{Witowski,Orlita}. The lower-frequency mode is
localized near graphene edge and propagates only in one direction determined by a magnetic field orientation, so it can
be guided along the edges and used to design plasmon circuits. Edge magnetoplasmons and possibilities of their
manipulation were extensively studied in semiconductor-based quantum Hall systems (see, e.g.,
\cite{Kumada2011,Andreev,Hashisaka} and references therein).

In several recent experiments \cite{Kumada,KumadaPRL,Petkovic,PetkovicPRL,KumadaNJP} the time-domain measurements of
edge magnetoplasmon propagation on graphene were carried out, which allowed to directly determine their velocities.
Similarly to that in semiconductor quantum wells \cite{Kumada2011,Andreev}, the velocity shows pronounced oscillations
as a function of the Landau level filling factor, decreasing at non-integer fillings where the system is conducting and
the dissipation is present. The presence of nearby metallic gate was also showed to reduce the plasmon velocity.
Although the general theory of magnetoplasmons \cite{Fetter,Volkov} allows to estimate their velocities, the effects of
screening and dissipation are insufficiently studied from the theoretical point of view. The existing approaches for
graphene magnetoplasmons \cite{Kumada,Wang,Petkovic} rely either on analytical formulas applicable for a clean system,
or use the Drude approximation for graphene conductivity, that cannot describe the oscillating filling-factor
dependencies originating from discreteness of Landau levels.

In this paper we provide the theoretical treatment of edge magnetoplasmons in graphene with taking into account gate
screening, dissipation and the filling-factor dependence of graphene optical conductivity in quantizing magnetic field.
In Sec.~\ref{Sec2} we consider the electromagnetic part of the problem solved using the Wiener-Hopf method and estimate
magnetoplasmon frequencies both in the absence and in the presence of dissipation. In contrast to conventional
calculations of a complex frequency accounting for the damping, we consider a real frequency and a complex wave vector.
In Sec.~\ref{Sec3} we calculate the conductivity tensor of disordered graphene in quantizing magnetic field using the
self-consistent Born approximation and length gauge which provide the qualitatively correct description of both its
low-frequency and filling-factor dependencies, both being crucial to the theory of edge magnetoplasmons. In
Sec.~\ref{Sec4} we show the results of numerical calculations of the magnetoplasmon dispersions and analyze how the
velocity depends on the Landau level filling factor and on the distance between graphene and metallic gate. In
agreement with the experiments \cite{Kumada,KumadaPRL,Petkovic,PetkovicPRL,KumadaNJP}, we find the oscillating behavior
of the velocity, which is suppressed by dissipation at non-integer Landau level fillings, and general reduction of the
velocity by the gate screening. Our conclusions are presented in Sec.~\ref{Sec5}.

\section{Electromagnetic problem}\label{Sec2}
\begin{figure}[t]
\begin{center}
\resizebox{0.85\columnwidth}{!}{\includegraphics{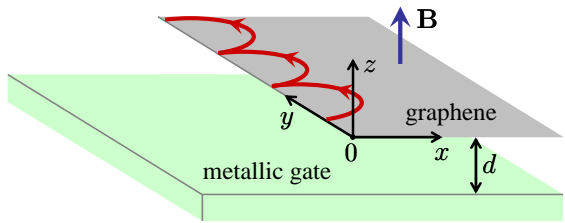}}
\end{center}
\caption{\label{Fig1}Trajectories of negatively charged electrons deflected by magnetic field $\mathbf{B}$ and bouncing
off the graphene edge, which participate in the formation of the magnetoplasmon mode propagating along the $y$ axis.
The nearby metallic gate at the distance $d$ is shown at the bottom.}
\end{figure}

Consider the magnetoplasmon wave propagating along the graphene edge, which is directed parallel to the $y$ axis, with
the frequency $\omega$ and wave vector $q$ (Fig.~\ref{Fig1}). To take into account the damping, we assume the complex
wave vector $\tilde{q}=q+i\alpha$ instead of a complex frequency. It is done in order to avoid complications arising in
many-body calculations of the retarded conductivities at complex frequencies and conforms the setups of time-domain
experiments where magnetoplasmon are attenuated in space. The graphene layer occupies the half-plane $x>0$, $z=0$. We
take its conductivity tensor $\sigma_{\alpha\beta}=\sigma_{\alpha\beta}(\omega)\Theta(x)$ ($\Theta$ is the unit step
function) in the long-wavelength limit $q\rightarrow0$ and assume it to be spatially uniform at $x>0$ and isotropic,
$\sigma_{xx}=\sigma_{yy}$, $\sigma_{xy}=-\sigma_{yx}$.

Writing the continuity equation $\partial\rho/\partial t+\mathrm{div}\,\mathbf{j}=0$ for the charge density
$\rho=\delta(z)e^{i(\tilde{q}y-\omega t)}\rho(x)$ and the currents $j_\alpha=-\sigma_{\alpha\beta}\nabla_\beta\varphi$
expressed in terms of the scalar potential $\varphi=e^{i(\tilde{q}y-\omega t)}\varphi(x,z)$, we obtain
\cite{Volkov,Wang}
\begin{eqnarray}
-i\omega\rho(x)=\Theta(x)\sigma_{xx}(\omega)(\partial_x^2-\tilde{q}^2)\varphi(x,0)\nonumber\\
+\delta(x)[\sigma_{xx}(\omega)\partial_x+i\tilde{q}\sigma_{xy}(\omega)]\varphi(x,0).\label{cont_eq}
\end{eqnarray}

The second equation arises from the three-dimensional Poisson equation $\varepsilon\nabla^2\varphi=-4\pi\rho$, where
$\varepsilon$ is the dielectric constant of the surrounding medium (we neglect the retardation in the low-frequency
limit). Doing a Fourier transform along the $x$ axis and taking into account that $\varphi|_{z=-d}=0$ due to the
grounded metallic gate, we can recast it into the integral equation \cite{Fetter,Volkov}
\begin{eqnarray}
\varphi(x,0)=\frac{4\pi}\varepsilon\int\limits_0^\infty dx'\:L_{\tilde{q},d}(x-x')\rho(x')\label{int_eq}
\end{eqnarray}
with the kernel
\begin{eqnarray}
L_{\tilde{q},d}(x)=\int\limits_{-\infty}^{+\infty}\frac{dk}{2\pi}
\frac{e^{ikx}}{\sqrt{k^2+\tilde{q}^2}\left[1+\coth(d\sqrt{k^2+\tilde{q}^2})\right]}.\label{kernel}
\end{eqnarray}

The system of equations (\ref{cont_eq})--(\ref{kernel}) determines the electromagnetic modes in the system. We can
solve it using the Wiener-Hopf method following Ref.~\cite{Volkov} with the only difference that we assume a real
frequency $\omega$ and a complex wave vector $\tilde{q}$ instead of complex $\omega$ and real $q$. The final equation
for edge magnetoplasmon dispersion is
\begin{eqnarray}
1+\frac{\sigma_{xy}(\omega)}{i\sigma_{xx}(\omega)}\tanh Z(\tilde{q},\omega)=0,\label{disp}
\end{eqnarray}
where
\begin{eqnarray}
Z(\tilde{q},\omega)=\frac1{2\pi}\int\limits_{-\infty}^{+\infty}\frac{\tilde{q}\:dk}{k^2+\tilde{q}^2}\nonumber\\
\times\ln\left\{1+\frac{4\pi
i\sigma_{xx}(\omega)}{\varepsilon\omega}\frac{\sqrt{k^2+\tilde{q}^2}}{1+\coth(d\sqrt{k^2+\tilde{q}^2})}\right\}.
\label{Z}
\end{eqnarray}
Here $\tilde{q}$ is assumed to have a positive real part, otherwise we need to replace it by $-\tilde{q}$ and change
the sigh of the second term in (\ref{disp}). If the dielectric constants above ($\varepsilon_1$) and below
($\varepsilon_2$) graphene layer are different, then $\varepsilon[1+\coth(d\sqrt{k^2+\tilde{q}^2})]$ in (\ref{Z}) is
replaced by $\varepsilon_1+\varepsilon_2\coth(d\sqrt{k^2+\tilde{q}^2})$, which complicates the following calculations
\cite{Volkov}. Nevertheless we can use the approximation of uniform medium with
$\varepsilon=(\varepsilon_1+\varepsilon_2)/2$ in the limit $d\rightarrow\infty$ and with $\varepsilon=\varepsilon_2$ at
$d\rightarrow0$.

Introducing the dimensionless quantity $\eta=4\pi\tilde{q}\sigma_{xx}(\omega)/i\omega$ \cite{Wang}, we can find $Z$ as
a function of $\eta$ and $\tilde{q}d$. We are interested in the long-wavelength and low-frequency limit, when
$\tilde{q},\omega\rightarrow0$ and $\eta\propto\sigma_{xx}(0)$. In the case of low dissipation we can assume
$|\eta|\ll1$, $\tilde{q}\approx q$, and calculate the analytical asymptotics \cite{Volkov} of $Z$ in this limit:
\begin{eqnarray}
Z\approx-\frac{\eta}{2\pi}\log\left(-\frac{4e}{\eta}\right),&qd\gtrsim e^{-\gamma_\mathrm{E}},\label{Z_asympt1}\\
Z\approx-\frac{\eta}{2\pi}\log\left(-\frac{4e^{1+\gamma_\mathrm{E}}qd}{\eta}\right),&
\displaystyle\frac{|\eta|}{4e^{\gamma_\mathrm{E}}}\lesssim qd\lesssim e^{-\gamma_\mathrm{E}},\label{Z_asympt2}\\
Z\approx\sqrt{-\eta qd},&\displaystyle qd\lesssim\frac{|\eta|}{4e^{\gamma_\mathrm{E}}},\label{Z_asympt3}
\end{eqnarray}
where $\gamma_\mathrm{E}\approx0.577$ is the Euler gamma constant. Eq.~(\ref{Z_asympt1}) corresponds to the case where
the gate is either absent or too far to influence the magnetoplasmons. Eq.~(\ref{Z_asympt2}) corresponds to the
opposite limit of local capacitance approximation \cite{Johnson}, when (\ref{int_eq}) reduces at small $d$ to the local
relationship for the plane capacitor: $\varphi(x)=4\pi d\rho(x)/\varepsilon$. Substituting
(\ref{Z_asympt1})--(\ref{Z_asympt3}) to (\ref{disp}), we obtain the dispersion relations for the edge magnetoplasmons
\cite{Volkov}:
\begin{eqnarray}
\omega=-\frac{2q\sigma_{xy}(0)}\varepsilon\ln\frac{2e}{qw},&qd\gtrsim e^{-\gamma_\mathrm{E}},\label{omega_clean1}\\
\omega=-\frac{2q\sigma_{xy}(0)}\varepsilon\ln\frac{2e^{1+\gamma_\mathrm{E}}d}w,&
\displaystyle\frac{qw}{2e^{\gamma_\mathrm{E}}}\lesssim qd\lesssim e^{-\gamma_\mathrm{E}},\label{omega_clean2}\\
\omega=-\frac{2\pi q\sigma_{xy}(0)}\varepsilon\sqrt{\frac{2d}w},& \displaystyle
qd\lesssim\frac{qw}{2e^{\gamma_\mathrm{E}}}, \label{omega_clean3}
\end{eqnarray}
where $w=-(2\pi/\varepsilon)[d\,\mathrm{Im}\,\sigma_{xx}(\omega)/d\omega]|_{\omega=0}$. Since $\sigma_{xy}(0)<0$ at
$\mathbf{B}\propto\mathbf{e}_z$, we have the edge mode propagating in the positive $y$ direction (Fig.~\ref{Fig1}).
Note that in vanishing magnetic field the dispersion equation (\ref{disp}) reduces to $Z(\tilde{q},\omega)=i\pi/2$
which typically implies $\eta$ of the order of unity ($\eta\approx2.4344$ at $d\rightarrow\infty$ \cite{Volkov}), so
the formulas (\ref{omega_clean1})--(\ref{omega_clean3}) derived under assumption $|\eta|\ll1$ become inapplicable.

In the presence of dissipation the expressions (\ref{omega_clean1})--(\ref{omega_clean3}) are inaccurate because in the
long-wavelength and low-frequency limit probed in the experiments
\cite{Kumada,KumadaPRL,Petkovic,PetkovicPRL,KumadaNJP} the damping rate dominates the frequency. This should happen
even in very clean graphene samples because the magnetoplasmon frequencies are far below the terahertz range. Therefore
the real part of $\sigma_{xx}$ dominates the imaginary part connected with $w$ (in other words, the dissipation
dominates electron inertial motion). Using (\ref{Z_asympt1}) and (\ref{Z_asympt3}) in (\ref{disp}) at
$\omega\rightarrow0$, we obtain the approximations for dispersion law and attenuation rate $\alpha$. At large
graphene-to-gate distance $qd\gtrsim1$ we obtain
\begin{eqnarray}
\omega=\frac{\pi q\sigma_{xx}(0)}{e\varepsilon\,\mathrm{Im}\,Y},\quad\alpha=-q\frac{\mathrm{Re}\,Y}{\mathrm{Im}\,Y},
\label{asympt1}
\end{eqnarray}
where $Y$ is the complex solution of the equation $1-iXY\ln Y=0$ with $\mathrm{Re}\,Y<0$, $\mathrm{Im}\,Y>0$, and
$X=-2e\sigma_{xy}(0)/\pi\sigma_{xx}(0)$. At very small distances, when $qd\lesssim0.01$, we obtain
\begin{eqnarray}
\omega=\frac{8\pi q^2\sigma_{xy}^2(0)d}{\varepsilon\sigma_{xx}(0)},\quad\alpha=q.\label{asympt3}
\end{eqnarray}
Note that, in contrast to the long-wavelength limit of the solution in Ref.~\cite{Johnson} with complex $\omega$ and
real $q$, where $\omega$ is purely imaginary, here we have the oscillations highly damped in space with
$\tilde{q}=q(1+i)$. At larger distances or wave vectors, when $qd\gtrsim0.01$, the dispersion becomes linear and
attenuation rate decreases.

\section{Optical conductivity in magnetic field}\label{Sec3}
To calculate the optical conductivity tensor $\sigma_{\alpha\beta}$ in disordered graphene in quantizing magnetic field
we use the version of the self-consistent Born approximation \cite{Shon} which allows us to take into account both
formation of Landau levels and their disorder-induced broadening. In this approximation the single-electron Green
functions are dressed by interaction with random disorder potential, which results in broadening of each Landau level,
and then the current vertex is modified by a disorder ladder. Direct application of the Kubo formula to calculate the
current response to the oscillating vector potential $\mathbf{A}=(c/i\omega)\mathbf{E}\propto e^{-i\omega t}$ provides
the dynamical conductivity
\begin{eqnarray}
\tilde\sigma_{\alpha\beta}(\mathbf{q},\omega)=\frac{ig}{\hbar\omega}G_{j_\alpha j_\beta}^\mathrm{R}(\mathbf{q},\omega)
\label{sigma_wrong}
\end{eqnarray}
in terms of the Fourier transform $G_{j_\alpha j_\beta}^\mathrm{R}(\mathbf{q},\omega)=-iS^{-1}\int
d\mathbf{r}d\mathbf{r}'\int_0^\infty dt\:e^{-i\mathbf{q}(\mathbf{r}-\mathbf{r}')+i\omega
t}\langle[j_\alpha(\mathbf{r},t),j_\beta(\mathbf{r}',0)]\rangle$ of the retarded Green function of the current
$j_\alpha(\mathbf{r},t)=ev_\mathrm{F}\psi^+(\mathbf{r},t)\sigma_\alpha\psi(\mathbf{r},t)$. Here $\psi(\mathbf{r},t)$ is
the two-component Heisenberg field operator of massless Dirac electrons in the valley $\mathbf{K}$ of graphene,
$v_\mathrm{F}\approx10^6\,\mbox{m/s}$ is the Fermi velocity, $g=4$ is the degeneracy over the valleys and spin
projections, $S$ is the system area.

However, in practical calculations of (\ref{sigma_wrong}) the problem of finite limit of $G_{j_\alpha
j_\beta}^\mathrm{R}(\mathbf{q},\omega)$ at $\omega\rightarrow0$ can be encountered (see \cite{Gusynin} and the example
of a clean system considered in Appendix~\ref{Appendix_A}). As result, $\tilde\sigma_{\alpha\beta}$ becomes divergent
at low frequencies which we are interested in, which is unphysical for a disordered conductor with nonzero dissipation.
At $q=0$, by definition of $G_{j_\alpha j_\beta}^\mathrm{R}$, this divergence corresponds to the unphysical response of
the current to static and uniform vector potential, that contradicts the gauge invariance. The same problem of spurious
response of graphene to vector potential, arising in calculations of graphene electromagnetic response functions where
the momentum cutoff was used to obtain finite results, was reported in Refs. \cite{Sabio,Principi,Takane}. The related
problem of spurious Meissner effect, appearing in the non-superconducting state of graphene when the superconducting
current of massless Dirac electrons is calculated, was reported \cite{Kopnin,Mizoguchi}. To overcome this problem, the
unphysical contribution to response functions can be calculated explicitly in the Dirac electron model and then
subtracted \cite{Principi}, the cutoff procedure can be modified \cite{Takane}, or auxiliary quadratic in momentum
terms can be added to Hamiltonian to make its spectrum bounded from below \cite{Mizoguchi}. Some of these approaches
can be modified and applied for electrons in graphene populating Landau levels in a clean system, as shown in
Appendix~\ref{Appendix_A}.

To calculate correct conductivity tensor, which is free from unphysical divergences in the low-frequency limit, we use
the $\mathbf{E}\cdot\mathbf{r}$ gauge, which does not involve such unobservable quantities as the vector potential from
the very beginning. The use of this gauge is justified in the dipole long-wavelength limit, which is applicable in our
case. Indeed, the experiments \cite{Kumada,KumadaPRL,Petkovic,PetkovicPRL,KumadaNJP} probe the plasmon wave vectors,
$q\ll l_H^{-1}$, much smaller than the inverse magnetic length $l_H=\sqrt{\hbar c/eB}$, therefore hereafter we consider
the limit $q\rightarrow0$ in conductivity calculations.  Using the spectral representation, we show in
Appendix~\ref{Appendix_B} that the optical conductivity in the $\mathbf{E}\cdot\mathbf{r}$ gauge is
\begin{eqnarray}
\sigma_{\alpha\beta}(\omega)=\frac{ig}{\hbar\omega}\left[G_{j_\alpha j_\beta}^\mathrm{R}(\omega)-G_{j_\alpha
j_\beta}^\mathrm{R}(\omega=0)\right].\label{sigma_correct}
\end{eqnarray}
In comparison with (\ref{sigma_wrong}), here the unphysical response $G_{j_\alpha j_\beta}^\mathrm{R}(\omega=0)$ is
subtracted and $\sigma_{\alpha\beta}(\omega)$ is guaranteed to be finite at $\omega\rightarrow0$. Thus the conductivity
calculation in the $\mathbf{E}\cdot\mathbf{r}$ gauge provides physically correct results satisfying the gauge
invariance. When Landau level widths are negligible, this approach reduces to those used in \cite{Principi,Takane}, as
shown by explicit calculations of $G_{j_\alpha j_\beta}^\mathrm{R}(\omega=0)$ in Appendix~\ref{Appendix_A}. Note that
the subtraction similar to those in (\ref{sigma_correct}) arises in the case of massive electrons due to the
diamagnetic contribution to conductivity \cite{Rammer}.

The self-consistent Born approximation provides the following expression for the Green function of currents (see also
\cite{Shon,Ando}):
\begin{eqnarray}
G_{j_\alpha j_\beta}^\mathrm{R}(\omega)=-\frac{e^2v_\mathrm{F}^2}{2\pi^2l_H^2}\sum_{n_1n_2}
\int\limits_{-\infty}^{+\infty}\frac{F_{\alpha\beta}^{n_1n_2}\:dz}{1-\gamma_{n_1n_2}G_{n_1}G_{n_2}^*}\nonumber\\
\times\left\{\frac{n_\mathrm{F}(z)G_{n_1}\,\mathrm{Im}\,G_{n_2}}{1-\gamma_{n_1n_2}G_{n_1}G_{n_2}}
+\frac{n_\mathrm{F}(z+\omega)\,\mathrm{Im}G_{n_1}\,G_{n_2}^*}{1-\gamma_{n_1n_2}G_{n_1}^*G_{n_2}^*}\right\},\label{G_SCBA}
\end{eqnarray}
where $G_{n_1}\equiv G_{n_1}^\mathrm{R}(z+\omega)$ and $G_{n_2}\equiv G_{n_2}^\mathrm{R}(z)$ are the retarded Green
functions of electrons on Landau levels with the numbers $n_1,n_2=0,\pm1,\pm2,\ldots$, $n_\mathrm{F}(z)=[e^{(\hbar
z-\mu)/T}+1]^{-1}$ is the Fermi-Dirac distribution, $\gamma_{n_1n_2}=(2\pi
l_H^2/\hbar^2S)\sum_{kk'}\langle\langle\psi_{n_1k'}|U|\psi_{n_1k}\rangle
\langle\psi_{n_2k}|U|\psi_{n_2k'}\rangle\rangle_{U}$ is the matrix element of the disorder potential $U$ averaged over
its realizations, which appears in the disorder ladder, $\psi_{nk}$ is the electron state on the $n$th Landau level
with the $k$th guiding center index. The factor $F_{\alpha\beta}^{n_1n_2}=2^{\delta_{n_10}+\delta_{n_20}-2}(c_\alpha
c_\beta^*\delta_{|n_1|-1,|n_2|}+c_\alpha^*c_\beta\delta_{|n_1|,|n_2|-1})$ with $c_x=1$, $c_y=i$ determines the
selection rules $|n_1|=|n_2|\pm1$ for the dipole inter-Landau level transitions in graphene.

Eq.~(\ref{G_SCBA}) takes the simple form in the case of short-range impurities, when $\gamma_{n_1n_2}$ vanish
\cite{Shon,Ando}. In this case we can take the electron Green functions $G_n^\mathrm{R}(\omega)=\int
d\omega'\:\rho_n(\omega')/(\omega-\omega'+i\delta)$ corresponding to the Lorentzian spectral density
$\rho_n(\omega)=(\hbar\Gamma/\pi)/[(\hbar\omega-E_n)^2+\Gamma^2]$, instead of half-elliptic densities \cite{Shon,Yang}
appearing as an artefact of the self-consistent Born approximation. Here $E_n=\mathrm{sgn}(n)\hbar
v_\mathrm{F}\sqrt{2|n|}/l_H$ and $\Gamma$ are, respectively, the energy and the width of the $n$th Landau level. We
assume equal widths of all Landau levels, in agreement with the scanning tunneling spectroscopy experiments (see, e.g.,
\cite{Miller}). From (\ref{sigma_correct})--(\ref{G_SCBA}) we obtain the final expression for the conductivity, which
behaves correctly in the $\omega\rightarrow0$ limit:
\begin{eqnarray}
\sigma_{\alpha\beta}(\omega)=\frac{ige^2v_\mathrm{F}^2}{2\pi\hbar l_H^2}\sum_{n_1n_2}F_{\alpha\beta}^{n_1n_2} \int
dz_1dz_2\:\rho_{n_1}(z_1)\rho_{n_2}(z_2)\nonumber\\
\times\frac{n_\mathrm{F}(z_2)-n_\mathrm{F}(z_1)}{(z_1-z_2)(\omega+z_2-z_1+i\delta)}.\label{sigma_Lorentz}
\end{eqnarray}

\begin{figure}[t]
\begin{center}
\resizebox{0.9\columnwidth}{!}{\includegraphics{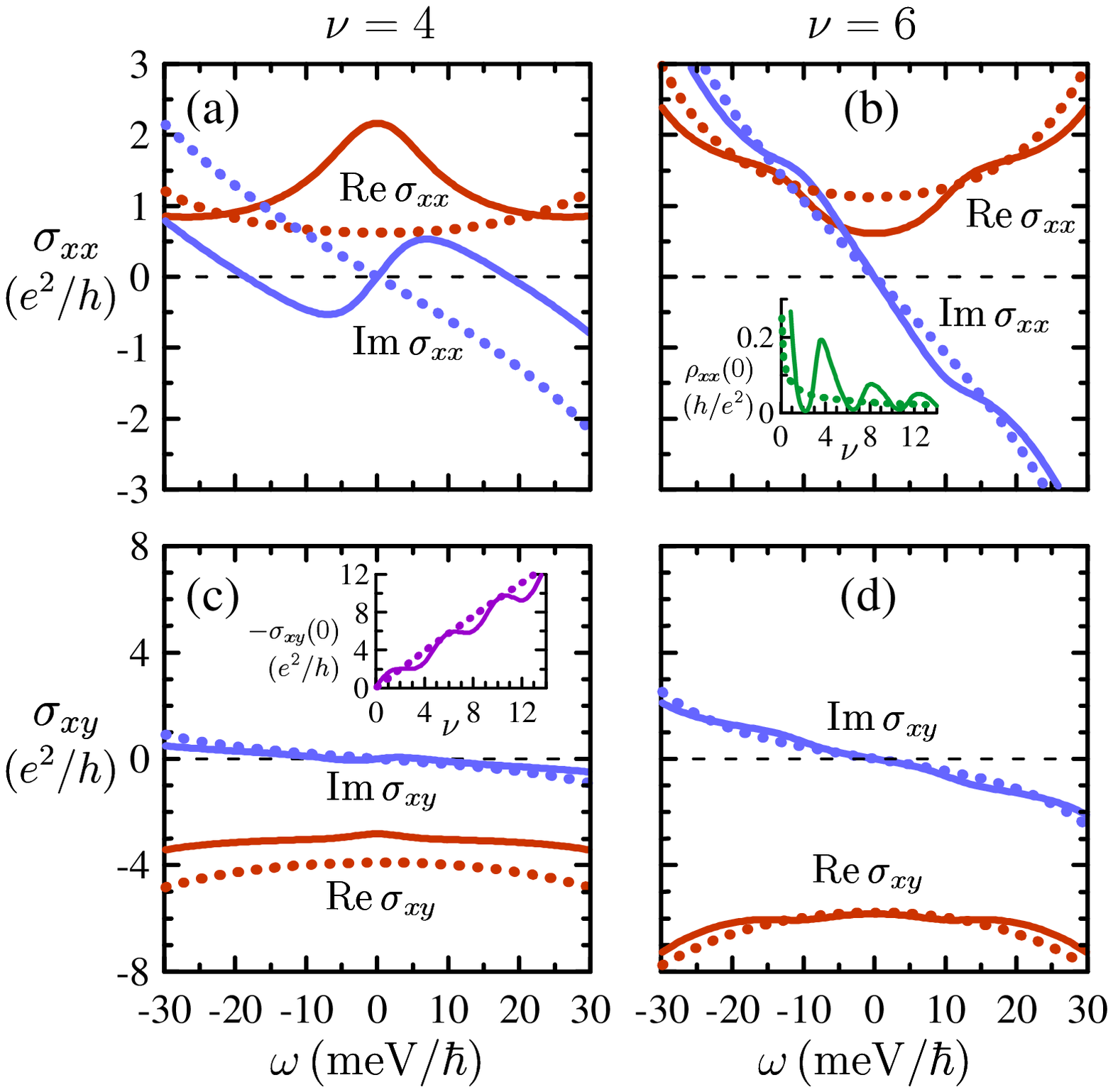}}
\end{center}
\caption{\label{Fig2}Solid lines: real and imaginary parts of $\sigma_{xx}$ (a,b) and $\sigma_{xy}$ (c,d) as functions
of $\omega$ calculated when the Fermi level is located in the center of the 1st Landau level, $\nu=4$ (a,c), or between
the 1st and 2nd levels, $\nu=6$ (b,d). Calculations are conducted at $B=12\,\mbox{T}$, $\Gamma=5\,\mbox{meV}$. Dotted
lines: Drude conductivities calculated at the same carrier density in the same magnetic field with
$\gamma=2\Gamma/\hbar$. Insets in (b) and (c) show the filling-factor dependencies of, respectively, static resistivity
$\rho_{xx}(0)$ and Hall conductivity $\sigma_{xy}(0)$ (solid lines) and their Drude counterparts (dotted lines).}
\end{figure}

The integrals in (\ref{sigma_Lorentz}) can be calculated analytically in the limit $T\rightarrow0$,
$n_\mathrm{F}(z)\rightarrow\Theta(\mu-\hbar z)$, corresponding to the experiments
\cite{Kumada,KumadaPRL,Petkovic,PetkovicPRL,KumadaNJP} carried out at cryogenic temperatures. The chemical potential
$\mu$ can be connected with the Landau level filling factor: $\nu=g\sum_n[\frac12\mathrm{sgn}(n)+\int
dz\:n_\mathrm{F}(z)\rho_n(z)]$; $\nu$ is zero for undoped graphene, where the 0th Landau level is half-filled, and
increases by 4 for each fully filled Landau level because of the fourfold degeneracy of electron states, so it equals
$4n+2$ when the $n$th level is completely filled and $4n$ when the $n$th level is half-filled. We do not take into
account the Zeeman splitting of Landau levels or possible symmetry breaking scenarios at fractional fillings because
they manifest themselves at much higher magnetic fields.

In the limit of high doping or low magnetic field, when the chemical potential $\mu$ is located between $E_n$ and
$E_{n+1}$, a single intraband transition $n\rightarrow n+1$ provides a major contribution to (\ref{sigma_Lorentz})
\cite{Wang,Witowski}. In the limit $n\gg1$ it takes the form of the classical Drude conductivity in magnetic field
\begin{eqnarray}
\sigma_{xx}(\omega)=\frac{n_\mathrm{c}e^2}{m^*}\frac{i(\omega+i\gamma)}
{(\omega+i\gamma)^2-\omega_\mathrm{c}^2},\label{sigmaxx_Drude}\\
\sigma_{xy}(\omega)=\frac{n_\mathrm{c}e^2}{m^*}\frac{\omega_\mathrm{c}}
{(\omega+i\gamma)^2-\omega_\mathrm{c}^2},\label{sigmaxy_Drude}
\end{eqnarray}
where $n_\mathrm{c}=\nu/2\pi l_H^2$ is the two-dimensional carrier density, $m^*=|\mu|/v_\mathrm{F}^2$ and
$\omega_\mathrm{c}=eB/m^*c$ are, respectively, the cyclotron mass and frequency, and $\gamma=2\Gamma/\hbar$ is the
decay rate of an electron-hole pair.

Examples of $\sigma_{\alpha\beta}(\omega)$ calculated from (\ref{sigma_Lorentz}) and in the Drude model
(\ref{sigmaxx_Drude})--(\ref{sigmaxy_Drude}) at the same carrier density are shown in Fig.~\ref{Fig2}. When the Fermi
level is located between Landau levels [Fig.~\ref{Fig2}(b,d)], the conductivity behavior at low frequencies is close to
the Drude model predictions, especially at high filling factors. At non-integer filling of Landau levels
[Fig.~\ref{Fig2}(a,c)], the conductivity deviates from the Drude model. Note the marked increase of $\sigma_{xx}$ at
low frequencies $|\omega|\lesssim\Gamma$ indicating the dissipation due to intralevel transitions. The second important
difference is the positive derivative $d\,\mathrm{Im}\,\sigma_{xx}(\omega)/d\omega|_{\omega=0}>0$ at the half-integer
filling in Fig.~\ref{Fig2}(a), which is indicative of effectively free electrons moving within the Landau level at
$|\omega|\lesssim\Gamma$. The similar behavior is demonstrated by a low-frequency Drude conductivity of a typical
conductor in the absence of magnetic field: $\mathrm{Im}\,\sigma(\omega)\propto\omega/(\omega^2+\gamma^2)\propto\omega$
at $\omega\rightarrow0$. In contrast, at the integer filling we see the negative derivative
$d\,\mathrm{Im}\,\sigma_{xx}(\omega)/d\omega|_{\omega=0}<0$ in Fig.~\ref{Fig2}(c), which is typical to bound electrons.
This derivative is related to the quantity $w$ in (\ref{omega_clean1})--(\ref{omega_clean3}), which can be interpreted
\cite{Volkov} as a distance where the energy of two-dimensional plasma oscillations is comparable with the cyclotron
energy. At half-integer fillings, when the dissipation is significant, this quantity has no such meaning, and the
formulas (\ref{omega_clean1})--(\ref{omega_clean3}) are inapplicable as well. Note the Drude conductivity
(\ref{sigmaxx_Drude}) always demonstrates the insulating behavior
$d\,\mathrm{Im}\,\sigma_{xx}(\omega)/d\omega|_{\omega=0}<0$ and thus cannot describe the transitions between conducting
and insulating regimes as $\nu$ is changed.

It is instructive to compare (\ref{sigma_Lorentz}) with the conductivity, which is initially calculated for a clean
graphene in magnetic field, and then supplemented with the phenomenological relaxation rate $2\Gamma/\hbar$
\cite{Gusynin}:
\begin{eqnarray}
\tilde{\tilde\sigma}_{\alpha\beta}(\omega)=\frac{ige^2\hbar v_\mathrm{F}^2}{2\pi
l_H^2}\sum_{n_1n_2}F_{\alpha\beta}^{n_1n_2}\nonumber\\
\times\frac{f_{n_2}-f_{n_1}}{(E_{n_1}-E_{n_2})(\hbar\omega+E_{n_2}-E_{n_1}+2i\Gamma)},\label{sigma_Phenom}
\end{eqnarray}
where $f_n=n_\mathrm{F}(E_n/\hbar)$. Despite the similarity between (\ref{sigma_Lorentz}) and (\ref{sigma_Phenom}), the
latter does not take into account electron transitions within the broadened Landau level, which are responsible for the
metallic-like behavior [Fig.~\ref{Fig2}(a)] of $\sigma_{\alpha\beta}(\omega)$ at non-integer Landau level fillings. At
low frequencies (\ref{sigma_Phenom}) is numerically very close to the Drude conductivity
(\ref{sigmaxx_Drude})--(\ref{sigmaxy_Drude}). Thus the phenomenological relaxation model, similarly to the Drude one,
cannot describe the sequence of insulating and conducting regimes.

Although our conductivity demonstrates the qualitatively correct low-frequency properties and properly takes into
account the Landau level quantization, further improvement is needed to achieve quantitative agreement with the
experiment. For example, the peaks in the static $\rho_{xx}(0)$ [see inset in Fig.~\ref{Fig2}(b)] and the
simultaneously occurring rising parts in the dependence of $\sigma_{xy}(0)$ on $\nu$ between the quantized plateaus
[inset in Fig.~\ref{Fig2}(c)] characteristic to conducting states at non-integer Landau level fillings are broader than
in the typical quantum Hall effect measurements \cite{Novoselov,Kumada}. From the other side, the broadening of these
peaks at nonzero frequencies studied in semiconductor quantum wells \cite{Engel,Saeed} should also been taken into
account. The conductivity model which explicitly includes consideration of localized and extended states would provide
more accurate results in the low-frequency region.

\section{Calculation results}\label{Sec4}
Using the formulas (\ref{disp}), (\ref{Z}), (\ref{sigma_Lorentz}), we can calculate numerically the dispersion
relations $\omega(q)$ and attenuation rates $\alpha(q)$ at different filling factors $\nu$ and graphene-to-gate
distances $d$ to study the effects of gate screening and dissipation. We take the parameters $B=12\,\mbox{T}$,
$\Gamma=5\,\mbox{meV}$, $\varepsilon=4$, which are close to the experimental conditions
\cite{Kumada,KumadaPRL,Petkovic,PetkovicPRL,KumadaNJP} where Landau quantization is well developed.

\begin{figure}[t]
\begin{center}
\resizebox{0.9\columnwidth}{!}{\includegraphics{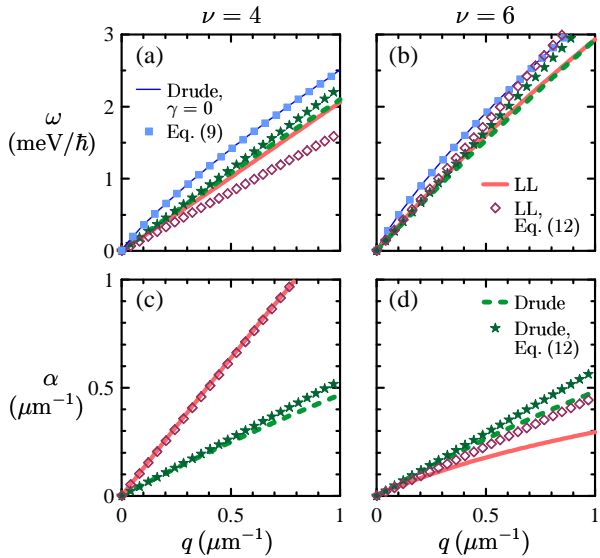}}
\end{center}
\caption{\label{Fig3}Edge magnetoplasmon frequencies $\omega$ (a,b) and attenuation rates $\alpha$ (c,d) as functions
of the wave vector $q$ calculated in the absence of gate screening at $d=\infty$, at half-integer [$\nu=4$, (a,c)] and
integer [$\nu=6$, (b,d)] Landau level fillings at $B=12\,\mbox{T}$, $\Gamma=5\,\mbox{meV}$, $\varepsilon=4$. The curves
show numerical calculation results with Landau-level based (LL) and Drude conductivities, and with Drude conductivity
in a clean ($\gamma=0$) system. The analytical approximations are shown by the points.}
\end{figure}

In Fig.~\ref{Fig3} we show typical examples of $\omega(q)$ and $\alpha(q)$ calculated in the absence of the gate
screening, at $d=\infty$, with the full Landau-level based conductivities (\ref{sigma_Lorentz}) and within the Drude
model (\ref{sigmaxx_Drude})--(\ref{sigmaxy_Drude}) at the same carrier density. At integer Landau level fillings
[$\nu=6$, Fig.~\ref{Fig3}(b,d)] $\omega$ and $\alpha$ are, respectively, slightly higher and significantly lower than
in the Drude model. This indicates that the dissipation, which suppresses $\omega$ and increases $\alpha$, is lower
than in the Drude model due to the inter-Landau level gap. At half-integer Landau level fillings [$\nu=4$,
Fig.~\ref{Fig3}(a,c)] the situation is opposite: the dissipation caused by the intralevel transitions slightly
suppresses $\omega$ and significantly increases $\alpha$ in comparison with the Drude model. Note also the pronounced
dissipation-induced decrease of $\omega$ and increase of $\alpha$ at $\nu=4$ in comparison with $\nu=6$. The numerical
calculations in both models are close to the analytical approximation (\ref{asympt1}) where the corresponding
conductivities at $\omega=0$ are substituted. For comparison we plotted the magnetoplasmon frequency calculated with
the Drude conductivity of a clean ($\gamma=0$) system, which is higher than in the disordered system and agrees with
the analytical approximation (\ref{omega_clean1}) very well.

\begin{figure}[t]
\begin{center}
\resizebox{0.9\columnwidth}{!}{\includegraphics{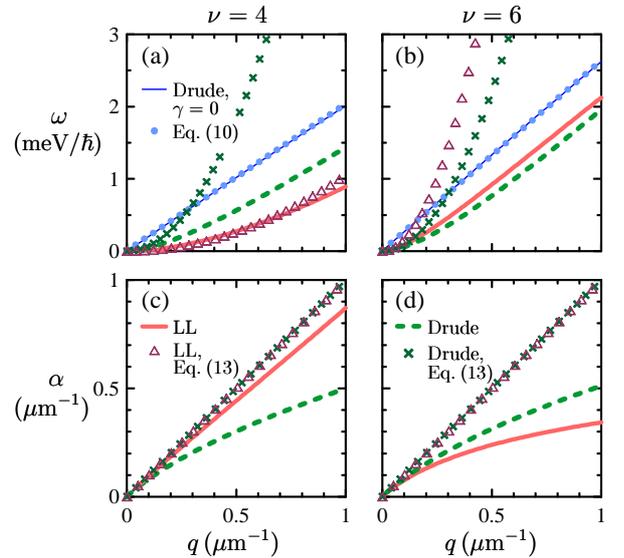}}
\end{center}
\caption{\label{Fig4}The same as in Fig.~\ref{Fig3} but in the presence of the gate screening at $d=200\,\mbox{nm}$.}
\end{figure}

The similar calculation results in the presence of the screening gate at $d=200\,\mbox{nm}$ are shown in
Fig.~\ref{Fig4}. We see the overall suppression of $\omega$ in comparison with the ungated case, which becomes even
stronger in the presence of dissipation. At integer [$\nu=6$, Fig.~\ref{Fig4}(b,d)] and half-integer [$\nu=4$,
Fig.~\ref{Fig4}(a,c)] Landau level fillings we again see the effect of, respectively, decreased and enhanced
dissipation on $\omega$ and $\alpha$. The analytical approximation (\ref{asympt3}) predicting highly damped mode with
quadratic dispersion at small $d$ is applicable only at very low distances or wave vectors, $qd\lesssim0.01$, while at
higher $q$ the dispersion laws become linear. Calculations with the Drude conductivity of a clean system agree with
(\ref{omega_clean3}) and provide considerably higher $\omega$.

\begin{figure}[t]
\begin{center}
\resizebox{0.9\columnwidth}{!}{\includegraphics{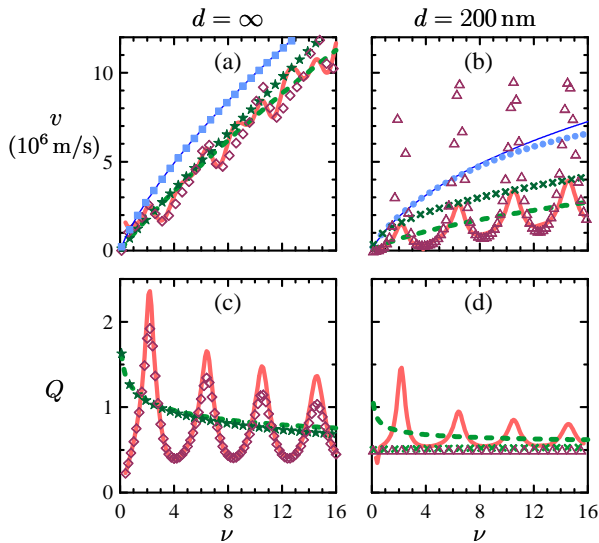}}
\end{center}
\caption{\label{Fig5} Edge magnetoplasmon velocities $v$ (a,b) and quality factors $Q$ (c,d) as functions of the Landau
level filling factor $\nu$ at $q=0.2\,\mu\mbox{m}^{-1}$ in the absence [$d=\infty$, (a,c)] and in the presence
[$d=200\,\mbox{nm}$, (b,d)] of the gate screening at $B=12\,\mbox{T}$, $\Gamma=5\,\mbox{meV}$, $\varepsilon=4$. The
curves show numerical calculation results with Landau-level based (LL) and Drude conductivities, and with Drude
conductivity in a clean system, and the points show analytical approximations. The notations for curves and points on
the panels (a,c) and (b,d) are the same as in, respectively, Figs.~\ref{Fig3} and \ref{Fig4}.}
\end{figure}

In Fig.~\ref{Fig5} we show the edge magnetoplasmon phase velocity $v=\omega/q$ and quality factor $Q=q/2\alpha$
calculated at $q=0.2\,\mu\mbox{m}^{-1}$ (where dispersions are almost linear) as functions of the filling factor $\nu$
both in the absence and in the presence of the metallic gate. As expected from the aforementioned dissipation-induced
frequency suppression, we observe the dips in both $v$ and $Q$ when the Fermi level is located in the centers of Landau
levels ($\nu=4,8,12,16$). In Fig.~\ref{Fig5}(a) these dips are slightly displaced to the left, perhaps, due to the
general rising trend of $v(\nu)$, and also demonstrate some extra oscillations, which can be an artefact of the model
used to calculate the conductivity. On the contrary, when the Fermi level is located in the middle of any inter-Landau
level gap ($\nu=2,6,10,14$), $v$ and $Q$ have peaks due to reduced dissipation. The oscillations of $v(\nu)$ and
$Q(\nu)$ occur around the smooth results of the quasiclassical Drude model, which is insensitive to how the individual
Landau levels are filled. Comparison with the calculations for a clean system shows that the gate screening not only
reduces the velocity, but also enhances the dissipation-induced suppression of $v$ and $Q$. As in the previous picture,
the analytical formula (\ref{asympt1}) well describes the dispersion and damping at $d=\infty$, while the formula
(\ref{asympt3}) for small $d$ is applicable only in the regions of low velocity and high dissipation.

\section{Conclusions}\label{Sec5}
We considered the magnetoplasmon modes propagating along graphene edge in quantizing magnetic fields in the presence of
the grounded metallic gate. The relationship between real frequencies and complex wave vectors (with the imaginary part
responsible for the damping) of these modes was obtained using the Wiener-Hopf method in a form of algebraic equation
(\ref{disp})--(\ref{Z}). The optical conductivity tensor $\sigma_{\alpha\beta}(\omega)$ was calculated at low
temperatures using the self-consistent Born approximation for graphene in magnetic field in the limit of short-range
impurities and with the assumption of Lorentzian broadening of Landau levels. The use of the
$\mathbf{E}\cdot\mathbf{r}$ gauge turned out to be convenient to calculate the conductivity which behaves correctly at
low frequencies. The quasiclassical Drude approximation to the conductivity, valid in the limit of large number of
filled Landau level, was also considered for comparison. The magnetoplasmon dispersions, attenuation rates, velocities
and quality factors were calculated both in the absence and in the presence of the gate at different Landau level
filling factors.

The analysis of the calculation results allows us to make the following conclusions:

(a) At integer filling of Landau levels ($\nu=2,6,10,\ldots$), where the Fermi level is located in the interlevel gap,
the dissipation is low, and the frequency, velocity and life time of the edge magnetoplasmon are increased in
comparison to the predictions of the Drude model. On the contrary, at half-integer filling of Landau levels
($\nu=4,8,12,\ldots$), where the Fermi level lies within a broadened Landau level, the dissipation is enhanced due to
intralevel transitions, and the frequency, velocity and life time of the edge magnetoplasmon are suppressed. So our
approach predicts the oscillations of the edge magnetoplasmon velocity when the filling factor is changed, which are
superimposed on the smooth trend of growing velocity as the carrier density increases. The commonly used
\cite{Wang,Witowski} Drude conductivity model describes only the latter because it does not take into account Landau
level quantization.

(b) The electric field screening caused by a nearby metallic gate decreases the frequency and velocity of the mode, and
makes their dissipation-induced suppression at non-integer filling of Landau levels much more pronounced.

(c) The analytical formulas (\ref{omega_clean1})--(\ref{omega_clean3}) for the mode frequency, which are frequently
used to analyze the experimental data \cite{Kumada,KumadaPRL,Petkovic,PetkovicPRL,KumadaNJP}, are applicable only for
very clean systems. The quantity $w$ entering these formulas lose its meaning at half-integer Landau level fillings,
when the dissipation is significant. Eq.~(\ref{asympt1}) can be used instead in the case when the gate screening is
negligible. The important feature of the experimental conditions is that plasmons are probed at relatively low
frequencies not exceeding the terahertz range, so any realistic rate of dissipation due to inter-Landau level
transitions dominates the frequency and substantially modifies mode properties in comparison with the predictions for a
clean system.

Our conclusions are in qualitative agreement with the experimental data
\cite{Kumada,KumadaPRL,Petkovic,PetkovicPRL,KumadaNJP}, where the suppression of the magnetoplasmon velocities at
non-integer Landau level fillings was observed. Our approach assume the abrupt edge of graphene and does not take into
account the details of spatial structure of Landau level wave functions near the edge, which include formation of
incompressible stripes \cite{Johnson}, edge channels \cite{Petkovic,Balev,Andreev}, and electron drift due to electric
field normal to the edge \cite{PetkovicPRL,KumadaPRL}. Analysis of these details requires an essential complication of
our approach and is beyond the scope of this paper.

To study the effects of Landau level filling on the edge magnetoplasmon we used the model for the optical conductivity
providing qualitatively correct description of its low-frequency behavior and the filling-factor dependence. A refined
model, which takes into account the localized states and their scaling properties, can provide an additional insight
into the theory of edge magnetoplasmons in quantum Hall regime and bring the results of our approach into better
quantitative agreement with the experiments.

\section*{Acknowledgments}
The authors are grateful to Oleg V. Kotov for helpful discussions. The work was supported by the grants No.
17-02-01134, 18-02-00985, and 18-52-00002 of the Russian Foundation for Basic Research. Yu.E.L. was partly supported by
the Program for Basic Research of the National Research University Higher School of Economics. A.A.S. acknowledges the
support from the Foundation for the Advancement of Theoretical Physics and Mathematics ``BASIS''.

\appendix

\section{Cutoff treatment in magnetic field}\label{Appendix_A}
In the clean limit the single-electron spectral densities are $\rho_n(\omega)=\delta(\omega-E_n/\hbar)$ and the Green
function of currents (\ref{G_SCBA}) at $\omega\rightarrow0$ takes the form (notations are taken from Sec.~\ref{Sec3}
and the limit $q\rightarrow0$ is implied)
\begin{eqnarray}
G_{j_\alpha j_\beta}^\mathrm{R}(0)=\frac{e^2\hbar v_\mathrm{F}^2}{2\pi
l_H^2}\sum_{n_1n_2}F_{\alpha\beta}^{n_1n_2}\frac{f_{n_2}-f_{n_1}}{E_{n_2}-E_{n_1}}.\label{G0_gen}
\end{eqnarray}
Imposing the usual cutoff for Landau level summation $-N\leqslant n_1,n_2\leqslant N$ and taking into account explicit
forms of $F_{\alpha\beta}^{n_1n_2}$ and $E_n$, we obtain after some algebra
\begin{eqnarray}
G_{j_\alpha j_\beta}^\mathrm{R}(0)=\frac{\delta_{\alpha\beta}e^2v_\mathrm{F}}{2\sqrt2\pi l_H}\nonumber\\
\times\sum_{n=0}^{N-1}\left[\sqrt{n+1}(f_{n+1}-f_{-n-1})-\sqrt{n}(f_n-f_{-n})\right].\label{G0_cutoff1}
\end{eqnarray}
Note that if even one of Landau level numbers $n_1$ or $n_2$ exceeds the cutoff then the whole transition
$n_2\rightarrow n_1$ is dropped out of the sum (\ref{G0_gen}). Since the cutoff energy $E_N$ should be much larger than
the temperature, we can assume $f_N=0$, $f_{-N}=1$, so after cancelation of the most of terms in (\ref{G0_cutoff1}) we
have
\begin{eqnarray}
G_{j_\alpha j_\beta}^\mathrm{R}(0)=-\frac{\delta_{\alpha\beta}e^2v_\mathrm{F}}{2\sqrt2\pi
l_H}\sqrt{N}=-\frac{\delta_{\alpha\beta}e^2E_N}{4\pi\hbar}.\label{G0_cutoff2}
\end{eqnarray}
Thus, in agreement with \cite{Sabio,Principi,Takane}, we obtain the unphysical response of graphene to static and
uniform vector potential proportional to the cutoff energy.

To overcome this problem, we can change the cutoff procedure by extending the summation over $n_1,n_2$ in
(\ref{G0_gen}) to infinity and transferring the cutoff into the occupation numbers by assuming
$f_n=n_\mathrm{F}(E_n/\hbar)\Theta(E_N-|E_n|)$ [with $\Theta(0)=1$]. The similar procedure for free electrons in the
absence of magnetic field was proposed in \cite{Takane}. Now if one of Landau level numbers in the transition
$n_2\rightarrow n_1$ exceeds the cutoff then the corresponding level is treated as unoccupied one, and the whole
transition is \emph{not} dropped out of the sum. The resulting response function with the modified cutoff procedure
\begin{eqnarray}
G_{j_\alpha j_\beta}^\mathrm{R}(0)=\frac{\delta_{\alpha\beta}e^2v_\mathrm{F}}{2\sqrt2\pi l_H}
\left\{\sum_{n=0}^{N-1}\sqrt{n+1}(f_{n+1}-f_{-n-1})\right.\nonumber\\
\left.-\sum_{n=0}^N\sqrt{n}(f_n-f_{-n})\right\}=0
\end{eqnarray}
vanishes in contrast to (\ref{G0_cutoff1}) due to different summation limits. However it is hard to apply this
procedure in a general case where interactions or disorder are present.

\section{Conductivity in the $\mathbf{E}\cdot\mathbf{r}$ gauge}\label{Appendix_B}
Calculating the conductivity as a response of the current $j_\alpha\equiv j_\alpha(\mathbf{q}=0)$ to the perturbation
of the Hamiltonian $-er_\beta E_\beta$ using the Kubo formula \cite{Giuliani}, we get
\begin{eqnarray}
\sigma_{\alpha\beta}(\omega)=\frac{ieg}{\hbar S}\int\limits_0^\infty dt\:e^{i(\omega+i\delta)t}
\left\langle[j_\alpha(t),r_\beta]\right\rangle.\label{sigma_length1}
\end{eqnarray}
Assuming existence of a complete set of many-body states $|n\rangle$ of the system with the energies $\tilde{E}_n$ and
occupation probabilities $p_n$, we find the spectral representation \cite{Giuliani} of (\ref{sigma_length1}):
\begin{eqnarray}
\sigma_{\alpha\beta}(\omega)=-\frac{eg}{\hbar S}\sum_{nk}(p_n-p_k)\frac{\langle n|j_\alpha|k\rangle\langle
k|r_\beta|n\rangle}{\omega-\omega_{kn}+i\delta},\label{sigma_length2}
\end{eqnarray}
where $\omega_{kn}=(\tilde{E}_k-\tilde{E}_n)/\hbar$. Using the commutation relation $[H,r_\alpha]=-i\hbar
v_\mathrm{F}\sigma_\alpha$ we obtain the relationship $\langle k|r_\beta|n\rangle=\langle
k|j_\beta|n\rangle/ie\omega_{kn}$, which allows to rewrite (\ref{sigma_length2}) as
\begin{eqnarray}
\sigma_{\alpha\beta}(\omega)=\frac{ig}{\hbar S}\sum_{nk}(p_n-p_k)\frac{\langle n|j_\alpha|k\rangle\langle
k|j_\beta|n\rangle}{\omega_{kn}(\omega-\omega_{kn}+i\delta)}.\label{sigma_length3}
\end{eqnarray}

On the other hand, we can write the spectral representation for the Green function of currents:
\begin{eqnarray}
G_{j_\alpha j_\beta}^\mathrm{R}(\omega)=\frac1S\sum_{nk}(p_n-p_k)\frac{\langle n|j_\alpha|k\rangle\langle
k|j_\beta|n\rangle}{\omega-\omega_{kn}+i\delta}.\label{G_spectral}
\end{eqnarray}
Using the simple equality
\begin{eqnarray}
\frac1{\omega-\omega_{kn}+i\delta}+\frac1{\omega_{kn}}=\frac\omega{\omega_{kn}(\omega-\omega_{kn}+i\delta)},
\end{eqnarray}
we finally obtain (\ref{sigma_correct}). Note that the subtraction of $G_{j_\alpha j_\beta}^\mathrm{R}(0)$, which
distinguishes (\ref{sigma_correct}) from (\ref{sigma_wrong}), turns out to be equivalent to replacement of $\omega$ in
the denominator to the frequency $\omega_{kn}$ of each individual transition in the sum. Sometimes this replacement is
made manually in approximate conductivity calculations in order to obtain physically reasonable results (see, e.g.,
\cite{Gusynin}).

The anomalous character of the nonzero $G_{j_\alpha j_\beta}^\mathrm{R}(0)$ can be revealed by taking $\omega=0$ in
(\ref{G_spectral}) and using again the relationship $\langle k|j_\beta|n\rangle=ie\omega_{kn}\langle
k|r_\beta|n\rangle$:
\begin{eqnarray}
G_{j_\alpha j_\beta}^\mathrm{R}(0)=\frac{ie}S\sum_{nk}p_n\left[\langle n|j_\alpha|k\rangle\langle
k|r_\beta|n\rangle\right.\nonumber\\
\left.-\langle n|r_\beta|k\rangle\langle k|j_\alpha|n\rangle\right].
\end{eqnarray}
If the summation over the states is made over complete basis, we can use the completeness condition
$\sum_k|k\rangle\langle k|=I$ and obtain $G_{j_\alpha j_\beta}^\mathrm{R}(0)=(ie/S)\langle[j_\alpha,r_\beta]\rangle=0$.
However the cutoff imposed in practical calculations prevents this cancelation as noted in \cite{Sabio} and illustrated
in Appendix~\ref{Appendix_A}.

\end{document}